\documentstyle[aps,twocolumn]{revtex}
\begin{document}
\author{Yong-Ping Zhang, Ying-Fang Gao, J. -Q. Liang}
\address{Institute of Theoretical Physics, Shanxi\\
University, Taiyuan, Shanxi, 030006, China}
\title{Measurement Induced Dephasing and Suppression of Persistent Current
in a Lattice Ring }
\maketitle

\begin{abstract}
We investigate the effect of measurement on the persistent current density
in a lattice ring penetrated by an Aharonov-Bohm flux and coupled to a
quantum dot which is continuously monitored by a point contact detector. It
is demonstrated by explicit analysis of the time-evolution of current
density that the persistent current in the lattice ring is suppressed
completely due to dephasing of quantum states induced by an analogous
which-way-measurment that whether or not the electron is localized on the
quantum dot. Practical experiments to observe the measurment induced
dephasing is proposed.
\end{abstract}


The properties of persistent currents in a mesoscopic ring induced by the
threading Aharonov-Bohm(AB)\ magnetic flux can be affected dramatically by
charge-transfer from the ring to a coupled quantum dot\cite{1} and such
phenomena have been well studied \cite{1,2,3,4,5,6,7} with an in-line or
side-branch quantum dot. B$\ddot{u}$ttiker and Stafford show in Ref.\cite{1}
that the persistent current is varied by the charge transfer from a side
branch quantum dot into a ring\ through a sequence of plateaus of
diamagnetic and paramagnetic states while exhibits sharp resonances with an
embedded quantum dot in the ring. Persistent current in the AB lattice ring
coupled weakly with a quantum dot is influenced by the spin fluctuation\cite%
{2} and{\bf \ }unusual features of persistent-current behaviors resulted by
the Kondo effect have been explored with a wide range of coupling constant%
\cite{2}. The lattice ring is described by an ideal one-dimensional
tight-binding model with $N$ lattice sites and the quantum dot is treated as
an Anderson impurity. The main interests of previous studies are focused on
the influence of Kondo effect on the persistent current which is considered
as the most convenient way to detect the Kondo screening cloud by experiments%
\cite{3,4,5,6,7} in which one can measure directly the current to
investigate the coherent transport through the quantum dot embedded in a
lattice ring. Based on perturbation analysis Affleck and Simon\cite{6} point
out that the persistent current depends strongly on the ratio of the
screening cloud size, which is the characteristic length scale associated
with the bulk Kondo temperature, to the ring circumference. The persistent
current is also a function of the total number of electrons and the magnetic
flux threading the ring besides the ratio\cite{7}.

Cedraschi, Ponomarenko and B$\ddot{u}$ttiker observe in Ref.\cite{8} an
interesting phenomenon that the persistent current in a mesoscopic ring is
sensitive to its environment, which is modeled by an embedded quantum dot
capacitively coupled to an external circuit with a dissipative impedance. It
is shown that the persistent current is strongly suppressed by zero-point
quantum fluctuation with decreasing external impedance\cite{8}. In the
present paper, we study the measuring effect on the ground state of
mesoscopic lattice ring coupled to a point contact through a side-branch
quantum dot (see Fig.1) which is regarded as a device to measure the
electron occupation of the dot. The point contact\cite{9,10} consists of two
reservoirs(emitter and collector) separated by an electrode as shown in
Fig.1. The mesoscopic ring coupled to a side-branch quantum dot which is
continuously monitored by the point contact. Electrons in the left and right
reservoirs fill up to different Fermi Levels, $\mu _{L}$, $\mu _{R}$,
respectively such that the electrons can flow through the electrode, for
example, from the left to right reservoirs. We assume that the quantum dot
has a single energy level $E_{0}$ and there are total $N$ sites in
mesoscopic lattice-ring. $V_{c}$ represents the coupling between neighboring
sites, and $V_{0}$ denotes coupling strength of the quantum dot with the
nearest site-1. The charge current flowing through the point contact can be
described by the well-known Landauer formula \cite{11} seen to be $I=\frac{eT%
}{2\pi }(u_{L}-u_{R})$, where $T$ is the transmission coefficient of the
contact. The current through the point contact is sensitive to nearby
electrostatic field and thus if an electron occupies the quantum dot, the
current decreases due to the electrostatic repulsion. Therefore the point
contact servers a measuring device to detect whether or not the quantum dot
is occupied by an electron. We show that the persistent current in the
mesoscopic ring will be strongly suppressed by the process to measure
electron occupation of the dot no matter how weak the coupling between the
point contact and the quantum dot is. In other words the quantum phase
coherence in the mesoscopic ring which gives rise to the persistent current
is completely destroyed by the continuous monitoring of the quantum dot by
the detector to determine if the electron exists on it ( a counterpart of
the which-way-measurement in the two slits interference experiment). 

The Hamiltonian describing the system is given by

\begin{equation}
H=H_{PC}+H_R+H_{int}.  \eqnum{1}
\end{equation}
where $H_{PC}$ represents the detector (quantum point contact)

\begin{equation}
H_{PC}=\sum_lE_la_l^{\dagger }a_l+\sum_rE_ra_r^{\dagger
}a_r+\sum_{l,r}\Omega _{lr}\left( a_r^{\dagger }a_l+h.c.\right) .  \eqnum{2}
\end{equation}
with $a_{l\left( r\right) }^{\dagger }$ and $a_{l\left( r\right) }$ being
the creation and annihilation operators in the left (right) reservoir, and $%
\Omega _{lr}$ denotes the hopping constant between the two reservoirs. The
Hamiltonian for the lattice ring coupled with a quantum dot of a single
energy level $E_0$ is

\begin{eqnarray}
H_R &=&E_0d_0^{\dagger }d_0+\sum_{i=1}^NE_id_i^{\dagger }d_i+V_c\left(
e^{i2\pi \frac \Phi {\Phi _0}}d_1^{\dagger }d_N+h.c.\right)  \nonumber \\
&&+\sum_{i=1}^{N-1}V_c\left( d_i^{\dagger }d_{i+1}+h.c.\right) +V_0\left(
d_0^{\dagger }d_1+h.c.\right) .  \eqnum{3}
\end{eqnarray}
where $d_0^{\dagger }$ $\left( d_0\right) $ is the creation (annihilation)
operator in the quantum dot and $d_i^{\dagger }$ $\left( d_i\right) $ is the
corresponding operator in the $i$-th lattice site $i=\{1,2,\cdots N\}$. The
magnetic flux $\Phi $ embraced by the lattice ring results in an AB phase $%
e^{i2\pi \frac \Phi {\Phi _0}}$ with $\Phi _0=\frac{hc}e$ being the quantum
unit of the flux, and $V_c$ represents hopping amplitude between the
adjacent sites, while the hopping constant between the dot and site-$1$ of
the lattice is denoted by $V_0$. The interaction Hamiltonian between the
point contact and the quantum dot in the case that the electron occupies the
quantum dot is seen to be

\begin{equation}
H_{int}=-\sum_{l,r}\delta \Omega _{lr}\left( a_r^{\dagger }a_l+h.c.\right)
d_0^{\dagger }d_0.  \eqnum{4}
\end{equation}
The transition amplitude between the two reservoirs of the point contact
decreases by a value of $\delta \Omega _{lr}$ due to the Comlomb repulsion
of electron occupying the dot. In the following we study the dynamics of the
quantum system with various initial conditions which indicate the result of
measurements at time $t=0$ and then evaluate the time-evolution of the
quantum state in order to determine the effect of measurements on the
current density in the ring. The time-dependent many-body state can be
constructed by all possible electron states as

\begin{eqnarray}
\left| \Psi \left( t\right) \right\rangle &=&[b_0\left( t\right)
d_0^{\dagger }+\sum_{i=1}^Nb_i\left( t\right) d_i^{\dagger
}+\sum_{l,r}b_{0lr}\left( t\right) d_0^{\dagger }a_r^{\dagger }a_l  \nonumber
\\
&&+\sum_{i=1}^Nb_{ilr}\left( t\right) d_i^{\dagger }a_r^{\dagger }a_l+\cdot
\cdot \cdot ]\left| 0\right\rangle .  \eqnum{5}
\end{eqnarray}
where, the vacuum state $\left| 0\right\rangle $ is defined such that there
is no electron in the coupled system (lattice ring and dot) and electrons
fill up to Fermi levels of the left and right reservoirs respectively.
Substitution of the state $\left| \Psi \left( t\right) \right\rangle $ into
time-dependent Schr$\ddot{o}$dinger equation, $i\frac d{dt}\left| \Psi
\left( t\right) \right\rangle =H$ $\left| \Psi \left( t\right) \right\rangle 
$, yields equations for the coefficients $b\left( t\right) $ which are the
probability amplitudes of the corresponding electron states. The density
matrices $\sigma _{ij}^{(n)}\left( t\right) $ are defined from these
amplitudes by tracing out the irrelative degrees of freedom with $n$ being
the number of electrons transmitting into the right reservoir of the point
contact at time $t$ and the total density-matrix is 
\begin{eqnarray}
\sigma _{ij}\left( t\right) &=&\sum_n\sigma _{ij}^{(n)}\left( t\right)
=b_i\left( t\right) b_j^{*}\left( t\right) +\sum_{l,r}b_{ilr}\left( t\right)
b_{jlr}^{*}\left( t\right)  \nonumber \\
&&+\sum_{l\succ l^{^{\prime }},r\succ r^{^{\prime }}}b_{ill^{^{\prime
}}rr^{^{\prime }}}\left( t\right) b_{jll^{^{\prime }}rr^{^{\prime
}}}^{*}\left( t\right) +\cdots .  \eqnum{6}
\end{eqnarray}
where, $i,j=\{0,1,2,\cdots N\}$ denote the occupation states of the quantum
dot for $i,j=0$ and the lattice sites for $i,j=1,2,\cdots N$ respectively.
The diagonal density-matrix element $\sigma _{ii}\left( t\right) $ indicates
the probability of electron occupying the state of the lattice site $i$,
while the off-diagonal element $\sigma _{ij(i\neq j)}\left( t\right) $
describes correlation between the states of the lattice sites $i$, $j$. The
density-matrix elements can be found from $b(t)$ with the help of the
Laplace transformation\cite{12} $b(E)=\int b(t)e^{iEt}dt$ . The sum over
energy levels of reservoirs which are assumed to be deeply inside the band ,
can be replaced by an integral, $\sum_{lr}\rightarrow \int \rho _l\rho
_rdE_ldE_r$ , with $\rho _l$, $\rho _r$ being the densities of states in the
emitter and collector (reservoirs) of the point contact. We furthermore
assume that the densities of states $\rho _l$, $\rho _r$, and the transition
amplitude $\Omega _{lr}$ are independent of energy\cite{12}. The rate
equations for density-matrix can be derived via the inverse Laplace
transformation and the result is \cite{9,10,12} , 
\begin{eqnarray}
\stackrel{.}{\sigma }_{0,0}^{(n)} &=&-D^{^{\prime }\ }\sigma
_{0,0}^{(n)}+D^{^{\prime }}\sigma _{0,0}^{(n-1)}+iV_0(\sigma
_{0,1}^{(n)}-\sigma _{1,0}^{(n)}),  \nonumber \\
\stackrel{.}{\sigma }_{1,1}^{(n)} &=&-D\sigma _{1,1}^{(n)}+D\sigma
_{1,1}^{(n-1)}-iV_0(\sigma _{0,1}^{(n)}-\sigma _{1,0}^{(n)})  \nonumber \\
&&+iV_c(e^{-i2\pi \frac \Phi {\Phi _0}}\sigma _{1,N}^{(n)}-e^{i2\pi \frac %
\Phi {\Phi _0}}\sigma _{N,1}^{(n)})  \nonumber \\
&&+iV_c(\sigma _{1,2}^{(n)}-\sigma _{2,1}^{(n)}),  \nonumber \\
\stackrel{.}{\sigma }_{i,i}^{(n)} &=&-D\sigma _{i,i}^{(n)}+D\sigma
_{i,i}^{(n-1)}+iV_c(\sigma _{i,i-1}^{(n)}-\sigma _{i-1,i}^{(n)})  \nonumber
\\
&&+iV_c(\sigma _{i,i+1}^{(n)}-\sigma _{i+1,i}^{(n)}),  \eqnum{7.a} \\
\stackrel{.}{\sigma }_{N,N}^{(n)} &=&-D\sigma _{N,N}^{(n)}+D\sigma
_{N,N}^{(n-1)}  \nonumber \\
&&+iV_c(\sigma _{N,N-1}^{(n)}-\sigma _{N-1,N}^{(n)})  \nonumber \\
&&-iV_c(e^{-i2\pi \frac \Phi {\Phi _0}}\sigma _{1,N}^{(n)}-e^{i2\pi \frac %
\Phi {\Phi _0}}\sigma _{N,1}^{(n)}),  \nonumber
\end{eqnarray}
for the diagonal elements and,

\begin{eqnarray}
\stackrel{.}{\sigma }_{0,1}^{(n)} &=&i\varepsilon _{1,0}\sigma _{0,1}^{(n)}-%
\frac 12(D+D^{^{\prime }\ })\sigma _{0,1}^{(n)}+\sqrt{DD^{^{\prime }\ }}%
\sigma _{0,1}^{(n-1)}  \nonumber \\
&&+iV_0(\sigma _{0,0}^{(n)}-\sigma _{1,1}^{(n)})+iV_c(\sigma
_{0,2}^{(n)}+e^{-i2\pi \frac \Phi {\Phi _0}}\sigma _{0,N}^{(n)}),  \nonumber
\\
\stackrel{.}{\sigma }_{0,i}^{(n)} &=&i\varepsilon _{i,0}\sigma _{0,i}^{(n)}-%
\frac 12(D+D^{^{\prime }\ })\sigma _{0,i}^{(n)}+\sqrt{DD^{^{\prime }\ }}%
\sigma _{0,i}^{(n-1)}  \nonumber \\
&&-iV_0\sigma _{1,i}^{(n)}+iV_c(\sigma _{0,i+1}^{(n)}+\sigma _{0,i-1}^{(n)}),
\nonumber \\
\stackrel{.}{\sigma }_{0,N}^{(n)} &=&i\varepsilon _{N,0}\sigma _{0,N}^{(n)}-%
\frac 12(D+D^{^{\prime }\ })\sigma _{0,N}^{(n)}+\sqrt{DD^{^{\prime }\ }}%
\sigma _{0,N}^{(n-1)}  \nonumber \\
&&-iV_0\sigma _{1,N}^{(n)}+iV_c(e^{i2\pi \frac \Phi {\Phi _0}}\sigma
_{0,1}^{(n)}+\sigma _{0,N-1}^{(n)}),  \nonumber \\
\stackrel{.}{\sigma }_{1,N}^{(n)} &=&i\varepsilon _{N,1}\sigma
_{1,N}^{(n)}-D\sigma _{1,N}^{(n)}+D\sigma _{1,N}^{(n-1)}  \eqnum{7.b} \\
&&+iV_c(\sigma _{1,N-1}^{(n)}-\sigma _{2,N}^{(n)})-iV_0\sigma _{0,N}^{(n)} 
\nonumber \\
&&+iV_ce^{i2\pi \frac \Phi {\Phi _0}}(\sigma _{1,1}^{(n)}-\sigma
_{N,N}^{(n)}),  \nonumber \\
\stackrel{.}{\sigma }_{1,i}^{(n)} &=&i\varepsilon _{i,1}\sigma
_{1,i}^{(n)}-iV_0\sigma _{0,i}^{(n)}-D\sigma _{1,i}^{(n)}+D\sigma
_{1,i}^{(n-1)}  \nonumber \\
&&+iV_c(\sigma _{1,i-1}^{(n)}+\sigma _{1,i+1}^{(n)}-\sigma
_{2,i}^{(n)}-e^{i2\pi \frac \Phi {\Phi _0}}\sigma _{N,i}^{(n)}),  \nonumber
\\
\stackrel{.}{\sigma }_{N,i}^{(n)} &=&i\varepsilon _{i,N}\sigma
_{N,i}^{(n)}-D\sigma _{N,i}^{(n)}+D\sigma _{N,i}^{(n-1)}  \nonumber \\
&&+iV_c(\sigma _{N,i-1}^{(n)}+\sigma _{N,i+1}^{(n)}-\sigma
_{N-1,i}^{(n)}-e^{-i2\pi \frac \Phi {\Phi _0}}\sigma _{1,i}^{(n)}), 
\nonumber \\
\stackrel{.}{\sigma }_{i,j}^{(n)} &=&i\varepsilon _{j,i}\sigma
_{i,j}^{(n)}-D\sigma _{i,j}^{(n)}+D\sigma _{i,j}^{(n-1)}  \nonumber \\
&&+iV_c(\sigma _{i,j-1}^{(n)}+\sigma _{i,j+1}^{(n)}-\sigma
_{i-1,j}^{(n)}-\sigma _{i+1,j}^{(n)}),  \nonumber \\
\stackrel{.}{\sigma }_{j,i}^{(n)} &=&(\stackrel{.}{\sigma }_{i,j}^{(n)})^{*}.
\nonumber
\end{eqnarray}
for the off-diagonal elements, where $\varepsilon _{i,j}=E_i-E_j$, $%
i,j=\{2,3\cdot \cdot \cdot N-1\}$ , $\Omega _{lr}^{^{\prime }}=\Omega
_{lr}-\delta \Omega _{lr}$ $D=2\pi \Omega _{lr}^2\rho _l\rho _r\left( \mu
_L-\mu _R\right) $, $D^{^{\prime }}=2\pi \Omega _{lr}^{^{\prime }2}\rho
_l\rho _r\left( \mu _L-\mu _R\right) $,and the overhead dot denotes the time
derivative. After the summation over $n$ (see Eq. $(6)$), the rate equations
are found in a compact form as 
\begin{eqnarray}
\frac{d\sigma _{ij}}{dt} &=&i\varepsilon _{ji}\sigma _{ij}+iV_{j-1,j}\sigma
_{i,j-1}-iV_{i,i-1}\sigma _{i-1,j}  \nonumber \\
&&+iV_{j+1,j}\sigma _{i,j+1}-iV_{i,i+1}\sigma _{i+1,j}  \nonumber \\
&&-iV_{1,N}\sigma _{N,j}\delta _{i,1}+iV_{N,1}\sigma _{i,N}\delta _{j,1} 
\eqnum{8} \\
&&-\frac \Gamma 2\sigma _{ij}\left( \delta _{i,0}+\delta _{j,0}\right)
(1-\delta _{i,j}).  \nonumber
\end{eqnarray}
where $\Gamma =(\sqrt{D}-\sqrt{D^{^{\prime }}})^2$ and $i,j=\left\{
0,1,\cdot \cdot \cdot N\right\} $. The boundary condition is obviously that $%
i,j=N+1=1$ and $V_{1,N}=V_{N,1}^{*}=V_ce^{i2\pi \frac \Phi {\Phi _0}}$ in
view of the AB phase. Furthermore, $V_{i,j}=V_0$ if $i$ or $j=0$ ; $%
V_{i,j}=0 $ if $i$ or $j<0$; otherwise $V_{i,j}=V_c$ .{\bf \ }From Eq. {\bf (%
}8{\bf )}, it is seen that the detector (the last term of Eq. (8{\bf )}\ )
has effect only on the off-diagonal element with the subscripts either $i$ or%
{\bf \ }$j$ being zero.\ For the case $\Gamma =0$\ i.e. $\delta \Omega
_{lr}=0$\ , Eq. (8) reduces to the rate equations for the lattice ring with
a side-coupled quantum dot discoupled from the detector. Based on the slave
boson mean field approach \cite{13}, the oscillating persistent current for
such a system has been found with the tight binding type Homiltionian in Ref.%
\cite{14}.

The time-dependent current density in the lattice-ring can be described by
the dimensionless quantity (measured in the unit $J_0=2eV_c/\hbar $)\cite{15}

\begin{equation}
J_{i,i+1}=%
%TCIMACRO{\func{Im} }%
%BeginExpansion
\mathop{\rm Im}
%EndExpansion
(\sigma _{i,i+1}).  \eqnum{9}
\end{equation}
which represents the current flows from the $i+1$ site to the $i$ site in
the lattice-ring. In stationary case, i.e. $\frac{d\sigma _{ij}}{dt}=0$,
Eqs. (8) possess a unique solution\cite{9} with vanishing off-diagonal
elements of the density matrix i.e.

\begin{equation}
\sigma _{i,j}(t\longrightarrow \infty )=0,(i\neq j).  \eqnum{10}
\end{equation}
which gives rise to vanishing current $J=0$ indicating the complete
dephasing due to the effect of the detector. {\bf \ }The physical
interpretation is obvious, since when the detector is discoupled with the
lattice-ring ($V_0=0$) the ground state of electron in the lattice ring is
in phase coherent state which leads to the persistent current induced by the
threading magnetic flux. This solution has been demonstrated by Cheung et al
in the Ref. [16]. Then after the coupling is turned on ($V_0\neq 0$), the
detector starts to monitor the coupled system and consequently destroys the
phase coherence of electron in the ring. The persistent current in the ring
vanishes at the asymptotic limit. To see this process clearly we in the
following provide the numerical solution of Eq.(8) in order to exhibit the
time-evolution of the off-diagonal elements of the density matrix. First of
all we consider the case that the quantum dot is not occupied by the
electron at time $t=0$, i.e.$\sigma _{0,0}(t=0)=0$ as measured by the
detector. Fig. 2 is a plot of the current density $J_{i,i+1}$ (here $i=1$
for example) via time $t$ with the initial condition that $\sigma
_{i,i}(t=0)=1/4$ for all $i\neq 0$, and the other elements are zero, i.e.
the electron occupies each site of ring with equal probability initially.
For the case that the detector is discoupled with the quantum dot ($\Gamma
=0 $) at $t>0$, the current density oscillates with time shown in Fig. $2a$
since the electron is periodically scattered by the quantum dot which is
considered as an impurity. While when the detector is turned on ( $\Gamma
\neq 0$ ) at $t>0$ the current oscillation damps to zero in a short time
period as shown in Fig. $2b$. The situation is similar as that shown in
Fig.2 for the case that the electron is located initially at any one site of
the ring. An opposite situation is that the detector has detected the
electron occupation of the quantum dot at time $t=0$. The initial condition
for this case is $\sigma _{0,0}(t=0)=1$, and $\sigma _{i,j}(t=0)=0$ for all $%
i,j\neq 0$. The numerical result shows that the time-evolution of the
current density $J_{1,2}(t)$ is again similar as in Fig.2 for both the
coupled and discoupled cases. Therefore the measurment induced suppression
of the persistent current in the lattice ring is independent of the initial
condition. It is also interesting to see the occupation probabilities of
electron on the quantum dot which are different dramatically for the two
cases of $\Gamma =0$\ and {\bf \ }$\Gamma \neq 0$ . In the former case the
occupation probability oscillates with time ( Fig. $3a$\ )and {\bf \ } in
the later case the oscillation of the occupation probability damps to a value%
{\bf \ }$\frac 15${\bf ( }Fig{\bf . }$3b$) due to the measuring.

We conclude that measurement of detector leads to the suppression of
persistent current in the mesoscopic ring. We emphasize that no matter how
weak the coupling between the detector and the quantum dot is, the
oscillation of current density damps to zero in the asymptotic limit. The
measuring localization of electron, namely whether or not the electron
exists at the quantum dot in our case, would destroy the phase coherence and
the persistent current completely. In this paper we display an analytic
evaluation of the measurement induced dephasing. The model described in Fig.
1 is, as a matter of fact, a practical set up of experiment in which the
current density is used as a probe to test the effect of measurement on the
quantum states.

{\bf Acknowledgment}: This work was supported by the Natural Science
Foundation of China under Grant No. 10475053

Figure Caption

Fig. 1. The mesoscopic ring coupled to a side-branch quantum dot which is
continuous monitored by the quantum point contact placed near the quantum
dot.

Fig.2 .The dimensionless current density $(J_{1,2})$ as a function of time
evaluated from the numerical solution of Eqs.(8) for $N=4$, $V_0=V_c=V$, $%
\varepsilon _{i,j}=0$, and $\Phi =0.1\Phi _0$ with the initial condition $%
\sigma _{i,i}(t=0)=1/4$, $\sigma _{0,0}(t=0)=0$ and $\sigma _{i,j}(t=0)=0$ ($%
i\neq j$). (a) The point contact is discoupled with the dot, $\Gamma =0$.
(b) The coupled case, $\Gamma =V$.

Fig. 3.The time-evolution of the electron-occupation-probability of the
quantum dot with initial condition $\sigma _{0,0}(t=0)=1$, and $\sigma
_{i,j}(t=0)=0$ for all $i,j\neq 0$. (a) $\sigma _{0,0}(t),$ for the
discoupled case $\Gamma =0$. (b) for the coupled case, $\Gamma =V$.


\bigskip

\begin{references}
\bibitem{1} M. B$\ddot{u}$ttiker and C. A. Stafford, Phys. Rev. Lett. 76,
495 (1996).

\bibitem{2} Sam Young Cho, Kicheon Kang, Chul Koo Kim, and Chang-Mo Ryu,
Phys. Rev. B 64, 033314 (2001).

\bibitem{3} K. Kang and S. -C. Shin, Phys. Rev. Lett. 85, 5619 (2000).

\bibitem{4} V. Ferrari $et$ $al.$, Phys. Rev. Lett. 82, 5088 (1999).

\bibitem{5} H. -P. Eckle, H. Johannesson, and C. A. Stafford, Phys. Rev.
Lett. 87, 016602 (2001).

\bibitem{6} I. Affleck and P. Simon, Phys. Rev. Lett. 86, 2854 (2001).

\bibitem{7} Hui Hu, Guang-Ming Zhang and Lu Yu, Phys. Rev. Lett. 86, 5558
(2001).

\bibitem{8} Pascal Cedraschi, Vadim V. Ponomarenko, and Markus B$\ddot{u}$%
ttiker, Phys. Rev. Lett. 84, 346 (2000).

\bibitem{9} S. A. Gurvitz, Phys. Rev. Lett. 85, 812 (2000).

\bibitem{10} S. A. Gurvitz, quant-ph/0212155; S. A. Gurvitz, Phys. Rev. B
56, 15215 (1997).

\bibitem{11} R. Landauer, J. Phys. Condens. Matter 1, 8099 (1989).

\bibitem{12} S. A.Gurvitz and Ya. S. Prager, Phys. Rev. B 53, 15932 (1996);
S. A. Gurvitz, Phys. Rev. B 57, 6602 (1998).

\bibitem{13} G. Kotliar and A. E. Ruckenstein, Phys. Rev. Lett. 57, 1362
(1986); V. Dorin and P. Schlottmann, Phys. Rev. B 47, 5095 (1993).

\bibitem{14} Guo-Hui Ding and Bing Dong, Phys. Rev. B 67, 195327 (2003).

\bibitem{15} P. A. Orellana, M. L. Ladr$\stackrel{^{\prime }}{o}$n de
Guevara, M. Pacheco, and A. Latg$\stackrel{^{\prime }}{e}$, Phys. Rev. B 68,
195321 (2003); P. A. Orellana, G. A. Lara, and Enrique V. \ Anda, Phys. Rev.
B 65, 155317 (2002).

\bibitem{16} Ho-Fai Cheung, Yuval Gefen, Eberhard K. Riedel, and Wei-Heng
Shih, Phys. Rev. B 37, 6050 (1988).
\end{references}
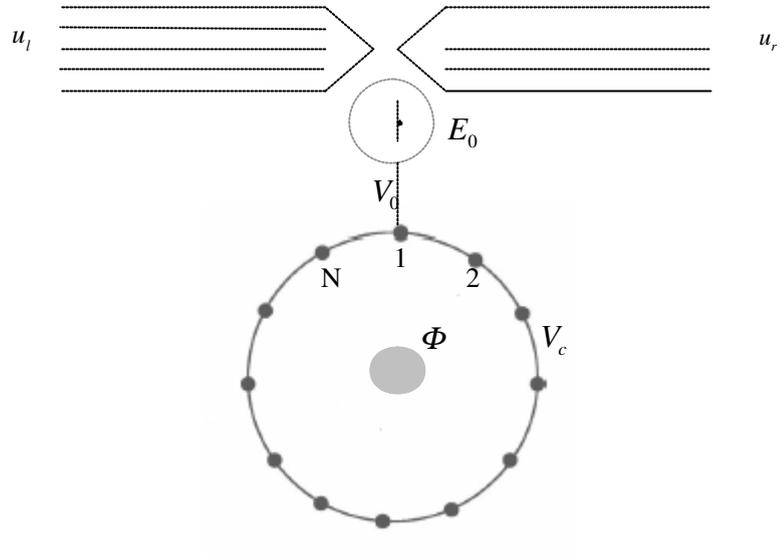
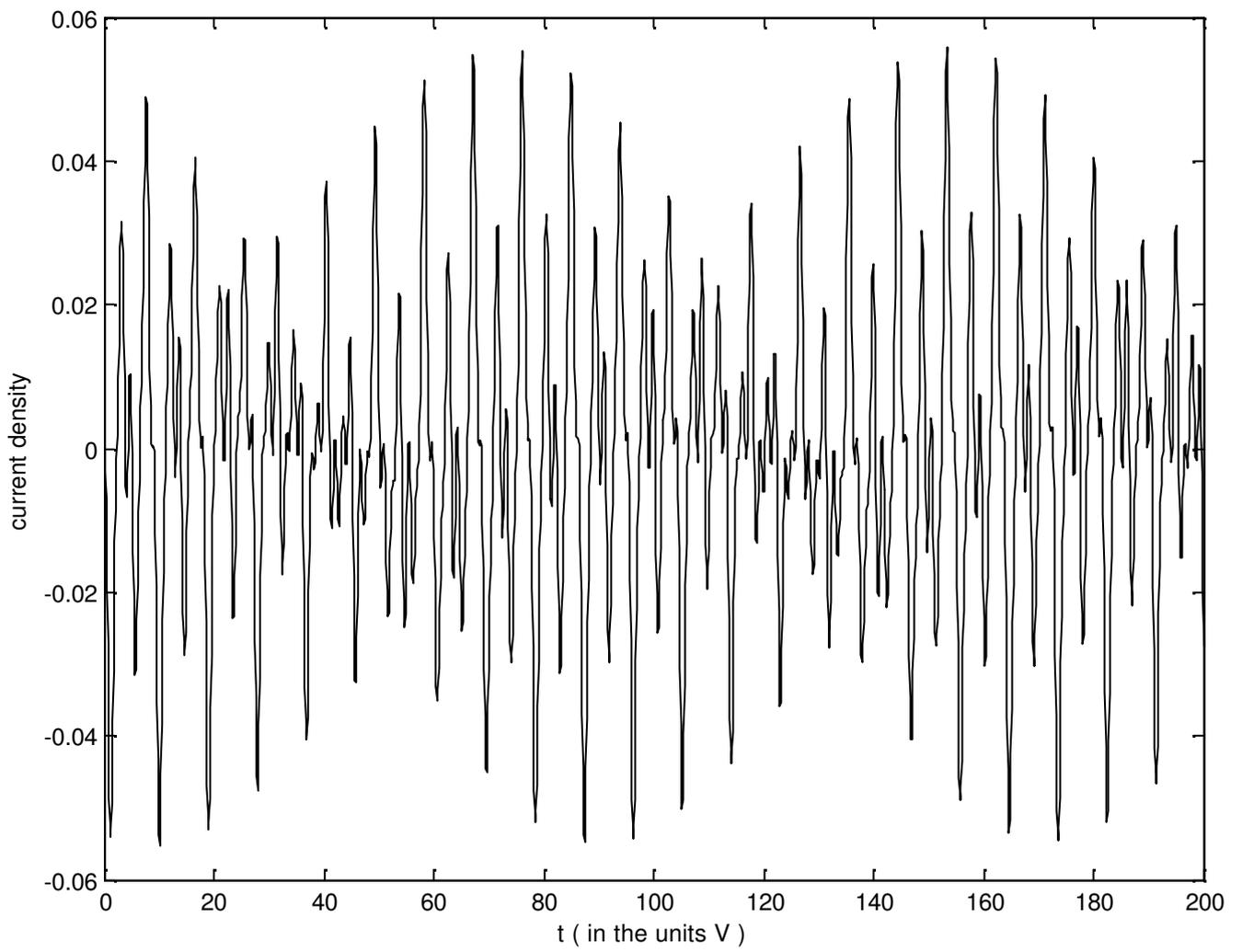
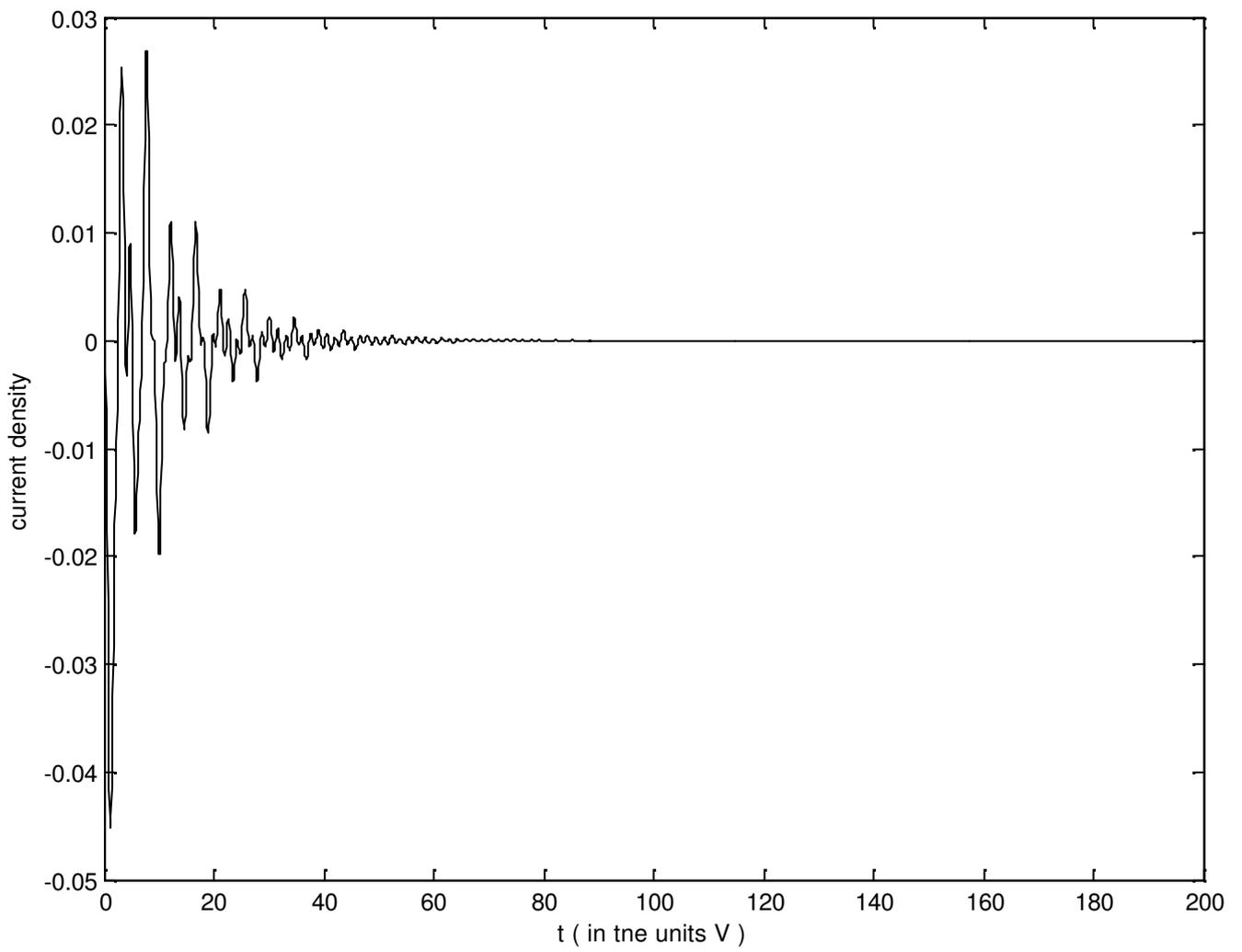
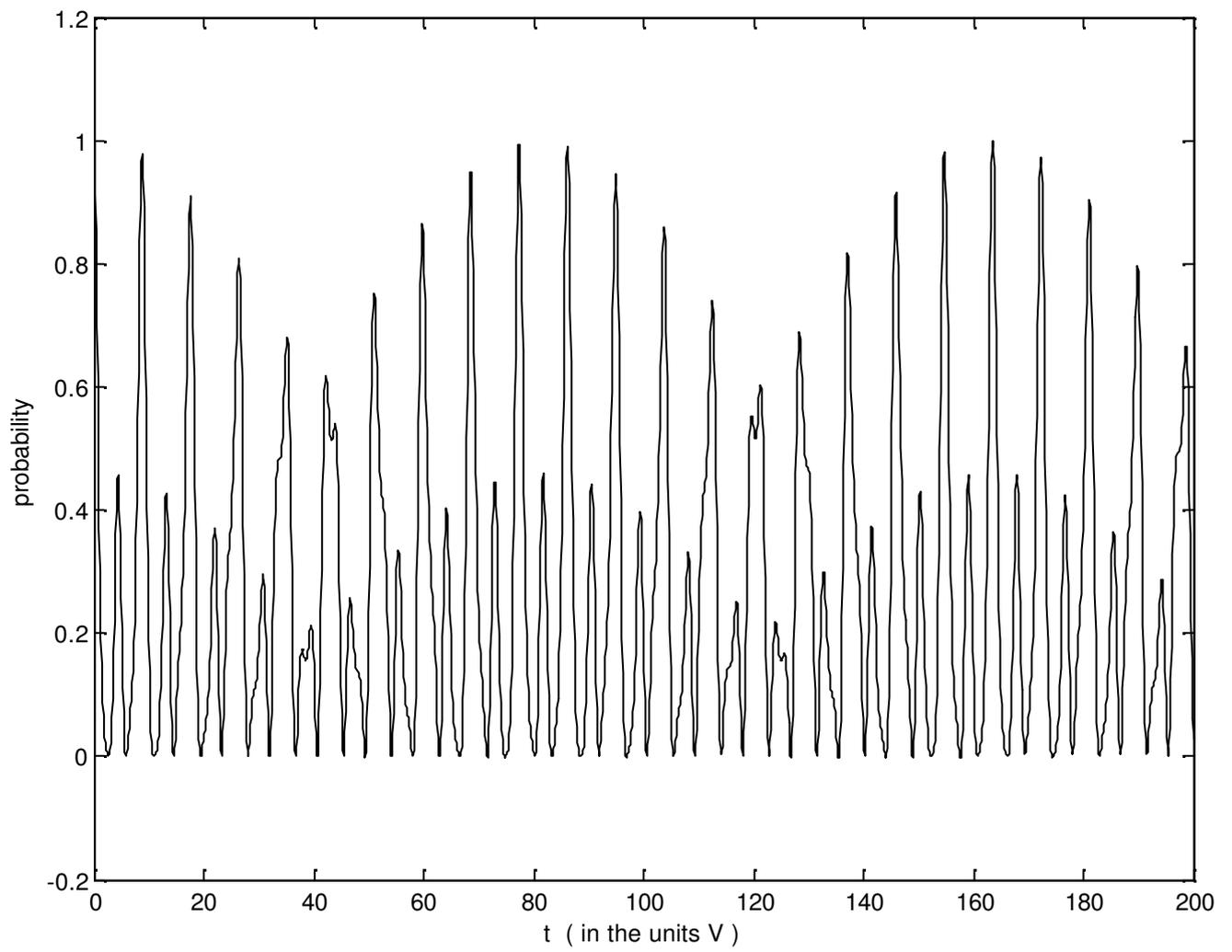
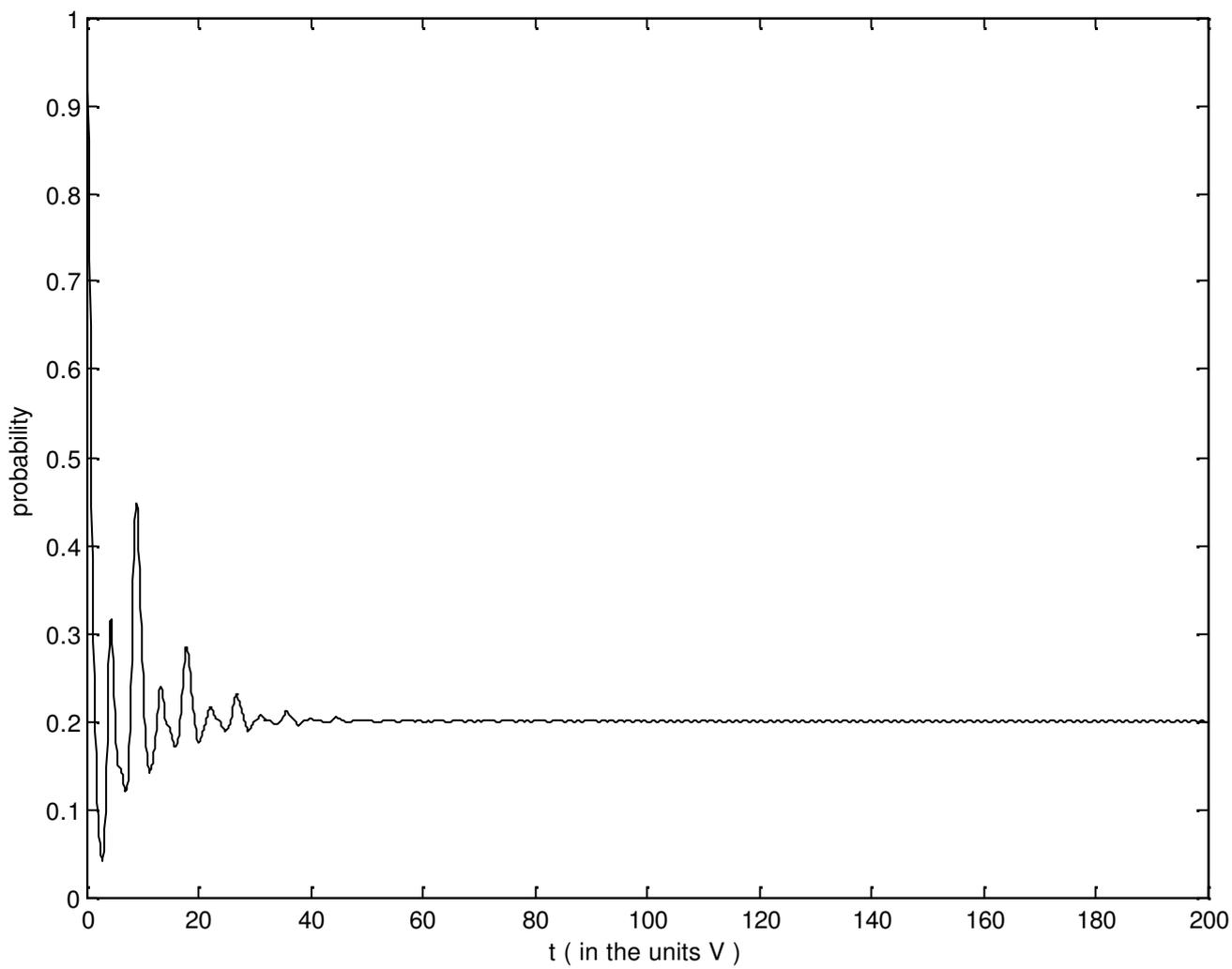
\end{document}